\documentclass[%
 aip,
apl,%
 amsmath,amssymb,
preprint,%
]{revtex4-1}

\usepackage{graphicx}% Include figure files
\usepackage{graphics}
\usepackage{dcolumn}% Align table columns on decimal point
\usepackage{bm}% bold math
\usepackage{upgreek}

\begin{document}

%\preprint{AIP/123-QED}

\title{Improving the read-out of the resonance frequency of nanotube mechanical resonators  \\}

\author{Jil Schwender}

\author{Ioannis Tsioutsios}
\altaffiliation[Current address: ]{Department of Applied Physics, Yale University, New Haven, CT 06520, USA.}

\author{Alexandros Tavernarakis}
\affiliation{ICFO-Institut de Ciencies Fotoniques, The Barcelona Institute of Science and Technology, 08860 Castelldefels, Barcelona, Spain}

\author{Quan Dong}
\affiliation{Centre de Nanosciences et de Nanotechnologies, CNRS, Univ. Paris-Sud, Univ. Paris-Saclay, C2N – Marcoussis, 91460 Marcoussis, France}

\author{Yong Jin}%
\affiliation{Centre de Nanosciences et de Nanotechnologies, CNRS, Univ. Paris-Sud, Univ. Paris-Saclay, C2N – Marcoussis, 91460 Marcoussis, France}

\author{Urs Staufer}
\affiliation{ICFO-Institut de Ciencies Fotoniques, The Barcelona Institute of Science and Technology, 08860 Castelldefels, Barcelona, Spain}

\affiliation{Department of Precision and Microsystems Engineering, Research Group of Micro and Nano-Engineering, Delft University of Technology, Mekelweg 2, 2628 CD, Delft, The Netherlands}

\author{Adrian Bachtold}
\email{adrian.bachtold@icfo.eu.}
\affiliation{ICFO-Institut de Ciencies Fotoniques, The Barcelona Institute of Science and Technology, 08860 Castelldefels, Barcelona, Spain}

\date{\today}
\begin{abstract}
We report on an electrical detection method of ultrasensitive carbon nanotube mechanical resonators. The noise floor of the detection method is reduced using a RLC resonator and an amplifier based on a high electron mobility transistor cooled at 4.2\,K. This allows us to resolve the resonance frequency of nanotube resonators with an unprecedented quality. We show that the noise of the resonance frequency measured at 4.2\,K is limited by the resonator itself, and not by the imprecision of the measurement. The Allan deviation reaches $\sim10^{-5}$ at 125\,ms integration time. When comparing the integration time dependence of the Allan deviation to a power law, the exponent approaches $\sim 1/4$. The Allan deviation might be limited by the diffusion of particles over the surface
of the nanotube. Our work holds promise for mass spectrometry and surface science experiments based on mechanical nano-resonators.
%
%Valid PACS numbers may be entered using the \verb+\pacs{#1}+ command.
\end{abstract}

%\pacs{Valid PACS appear here}% PACS, the Physics and Astronomy
                             % Classification Scheme.
%\keywords{Suggested keywords}%Use showkeys class option if keyword
                              %display desired
\maketitle

In recent years mechanical nano-resonators have been proven to be exceptional sensors of external forces \cite{mamin_sub-attonewton_2001,gavartin_hybrid_2012,moser_ultrasensitive_2013,heritier_nanoladder_2018} and adsorption of mass \cite{yang_zeptogram-scale_2006,chiu_atomic-scale_2008,chaste_nanomechanical_2012}. Mechanical resonators can be used for scanning probe microscopy \cite{rossi_vectorial_2017,de_lepinay_universal_2017}, magnetic resonance imaging \cite{rugar_single_2004,rugar_single_2004,degen_nanoscale_2009}, mass spectrometry \cite{hanay_single-protein_2012} and the study of surface science \cite{wang_phase_2010,yang_surface_2011,tavernarakis_atomic_2014}. This includes the diffusion of adsorbed atoms on the surface of a resonator \cite{yang_surface_2011,atalaya_diffusion-induced_2011,rhen_particle_2016}, the formation of monolayers of adsorbed atoms in the solid and liquid phases \cite{tavernarakis_atomic_2014}, and phase transitions \cite{wang_phase_2010}. Key for all these studies is a low noise transduction of the mechanical motion \cite{stapfner_cavity-enhanced_2013,singh_optomechanical_2014,song_graphene_2014,cole_evanescent-field_2015,guttinger_energy-dependent_2017,tavernarakis_optomechanics_2018}.

Resonators made out of carbon nanotubes hold the records in force and mass sensitivities thanks to their incredible small masses. However, the transduction of the nanotube vibrations into a measurable signal is a challenging task. Because nanotubes are small, the transduced signal is minuscule. Moreover, the read-out scheme has to be compatible with the low-temperature setups used to achieve the highest sensitivities in force \cite{moser_ultrasensitive_2013} and mass \cite{chaste_nanomechanical_2012} detection. Furthermore, electrical transduction schemes are limited by parasitic capacitances from the device pads and cables to the ground, setting a cut-off limit for the read-out frequency typically in the range between 1 and 10\,kHz. Efficient read-out was demonstrated by downmixing the high frequency signal of the motion to a low-frequency (1-10\,kHz) current modulation using the nanotube as a mixer. At these relatively low frequencies, the current suffers from large 1/f noise. Moreover, the current amplification took place at room temperature, so that the amplifier noise and the parasitic noise picked up by the cable between the device and the amplifier significantly contribute to the total noise.

In this letter, we develop an electrical downmixing read-out of nanomechanical motion with reduced noise compared to previous works. It operates at higher frequencies ($\sim$ 1.6\,MHz) using an RLC resonator. The current amplification is done with a high electron mobility transistor (HEMT) at liquid helium temperature \cite{dong_ultra-low_2014}. We demonstrate an improved detection of the resonance frequency of carbon nanotube resonators. The frequency stability is no longer limited by additive noise related to the imprecision of the detection, but by noise intrinsic to the device.

\begin{figure}
\includegraphics{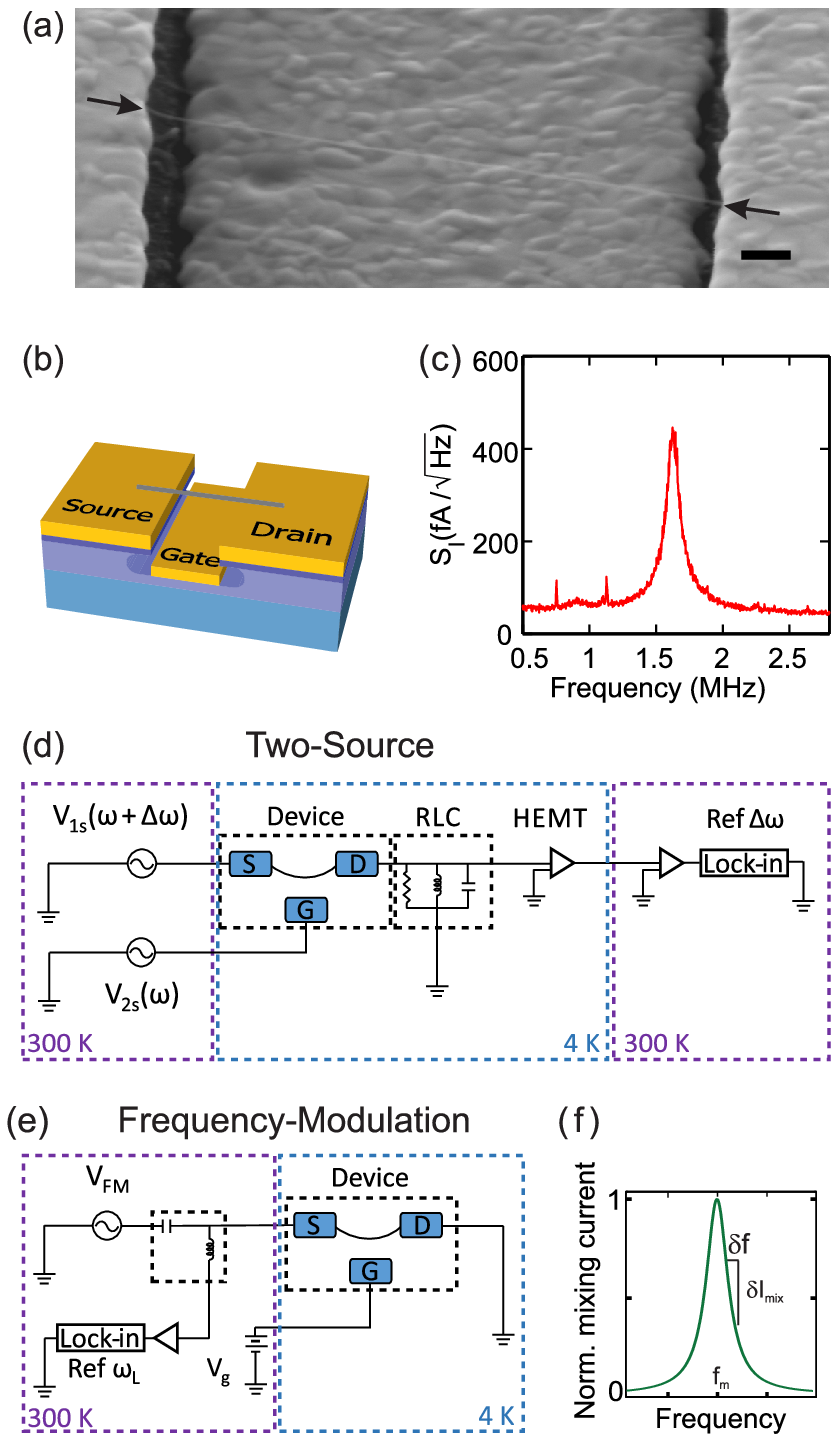}
\caption{\label{fig:fig_1} (a) Scanning electron microscopy image of a typical nanotube resonator. The nanotube is marked by black arrows. The scale bar is 100 nm. (b) Sketch of the device. The nanotube is contacted electrically to two metal electrodes and suspended over a local gate electrode. The length of the suspended part of the nanotube is $\sim$1\,$\mathrm{\mu}$m  and the distance to the gate is $\sim$350\,nm. (c) Current noise spectrum of the RLC resonator at 4.2\,K. (d) Two-source setup. We apply an oscillating voltage with amplitude V$_{\mathrm{1s}}$ at frequency $\omega+\Delta\omega$ onto the source electrode and an oscillating voltage with amplitude V$_{\mathrm{2s}}$ at frequency $\omega$ on the gate electrode.  We set $\Delta\omega$ equal to the RLC resonance frequency. (e) FM setup. We drive and detect the nanotube vibration in reflection by applying a frequency-modulated oscillating voltage with amplitude V$_\mathrm{FM}$ onto the source electrode. (f) Current fluctuations $\delta I_{\mathrm{mix}}$ are related to frequency fluctuations $\delta f$  via the slope of the resonance line shape.}
\end{figure}

All the measurements were carried out in a homebuilt ultra-high vacuum cryostat \cite{chaste_nanomechanical_2012} at a base pressure of 3$\cdot10^{-11}$\,mbar and 4.2\,K. The measured device consists of a nanotube contacted by two electrodes and suspended over a local gate electrode as shown in Figs. \ref{fig:fig_1}(a,b). The carbon nanotube was grown by chemical vapor deposition on a substrate containing prepatterned electrodes in the last step of the fabrication process in order to diminish contamination \cite{eichler_nonlinear_2011}. The mechanical motion was driven and detected using two different methods as shown in Figs. \ref{fig:fig_1}(d,e), often called two-source mixing \cite{sazonova_tunable_2004} and frequency-modulation (FM) mixing \cite{gouttenoire_digital_2010}. The two-source mixing was used together with the low noise read-out setup consisting of the RLC resonator and the HEMT based on an AlGaAs/GaAs heterostructure. The FM mixing is not compatible with the high-frequency of the RLC resonator, but we used it as a benchmark, as it has enabled the best frequency stability measurements of nanotube resonators thus far \cite{chaste_nanomechanical_2012}.

Both measurement techniques rely on the nanotube-gate capacitance oscillation generated by the motion of the resonator, which in turn modulates the measured current \cite{knobel_nanometre-scale_2003}. The two-source method generates a current directly proportional to the amplitude of mechanical vibrations. By contrast, the current in the FM method is related to the derivative of the real part of the response of the resonator \cite{gouttenoire_digital_2010}. In both methods, we measured the noise of the resonance frequency by driving the resonator at a setpoint frequency where the slope of the response is highest (Fig. \ref{fig:fig_1}(f)). A change in resonance frequency leads to a change in current. When the resonance frequency drifts more than a certain limit value, we use a computer-controlled feedback loop to correct the driving frequency in order to return to the setpoint for which the slope of the response is highest.

Our measurement scheme features a remarkably low current noise floor (Fig. \ref{fig:fig_1}(c)). The total noise floor is 0.43\,pA/$\sqrt{\mathrm{Hz}}$; it is estimated from the noise at the RLC resonance frequency. For comparison, the noise floor of the FM mixing is about 16\,pA/$\sqrt{\mathrm{Hz}}$. The two main contributions to this low noise floor are the noise picked up at the level of the sample copper box, which is left partially open in order to be able to evaporate atoms onto the nanotube, and the Johnson-Nyquist noise of the impedance of the RLC resonator (0.18\,pA/$\sqrt{\mathrm{Hz}}$). The inductance of the circuit is given by the 33\,$\mathrm{\mu}$H inductance soldered onto a printed-circuit board (PCB). The 290\,pF capacitance measured from the RLC resonance frequency comes from the 200\,pF capacitance soldered on the PCB and the capacitance of the radio-frequency cables between the device and the HEMT. The 6.66\,k$\Omega$ resistance obtained from the RLC line-width is attributed to the 10\,k$\Omega$ resistance soldered onto the PCB and the input impedance of the HEMT. The gain of the HEMT is 2.6; it is estimated from the temperature dependence of the Johnson-Nyquist noise. The signal is amplified at room temperature by the amplifier SA-220F5 protected in a copper box.

The Allan deviation of the resonance frequency provides information on the nature of the noise of the mechanical resonator. We compute the Allan deviation from the measured time traces of the resonance frequency as \cite{allan_time_1987}
\begin{equation}
 \sigma_{\mathrm{Allan}}( \tau_{\mathrm{A}})=  \sqrt{\frac{1}{2(N-1)}\sum_{i=1}^{N-1}\left(\frac{\overline{f}_{i+1}-\overline{f}_{i}}{f_{0}}\right)^{2}},
\label{eq:Allan}
\end{equation}
with $\overline f_{i}$ being the averaged frequency during the time interval $i$ of length $\tau_{\mathrm{A}}$, $f_{0}$ the resonance frequency averaged over the whole measurement and $N$ the total number of time intervals.
The Allan deviation is usually plotted as a function of $\tau_{\mathrm{A}}$, which is often called the integration time. The Allan deviation can be seen as the time-domain equivalent of the power spectral density of the noise of the resonance frequency. When the frequency noise is a 1/f noise, the Allan deviation remains constant when increasing $\tau_{\mathrm{A}}$.  When the frequency noise is white, the Allan deviation scales as $1/\sqrt{\tau_{\mathrm{A}}}$  \, \cite{cleland_noise_2002}.

Imprecision noise in the detection of the vibrations also contributes to the Allan deviation. The imprecision noise of the detection and the noise of the resonance frequency both contribute to the measured current noise (Fig. \ref{fig:fig_1}(f)), so that they cannot be distinguished. The contribution of detection imprecision to the Allan deviation is given by

\begin{equation}
\sigma_{\mathrm{Allan}}(\tau_{\mathrm{A}}) \simeq  \frac{\Delta f}{f_{0}} \frac{N_{\mathrm{T}}}{S} \sqrt{\frac{1}{2 \pi \tau_{\mathrm{A}}}},
\label{eq:Allan_add_phase_noise}
\end{equation}
where $S$ is the current amplitude of the driven resonance at $f_0$, $\Delta f$ the full width at half maximum of the resonance, and $N_{\mathrm{T}}$ the current noise floor discussed above. Here $1/2\pi\tau_{\mathrm{A}}$ is the measurement bandwidth with a first-order low pass filter \cite{sansa_frequency_2016}.

\begin{figure}
\includegraphics{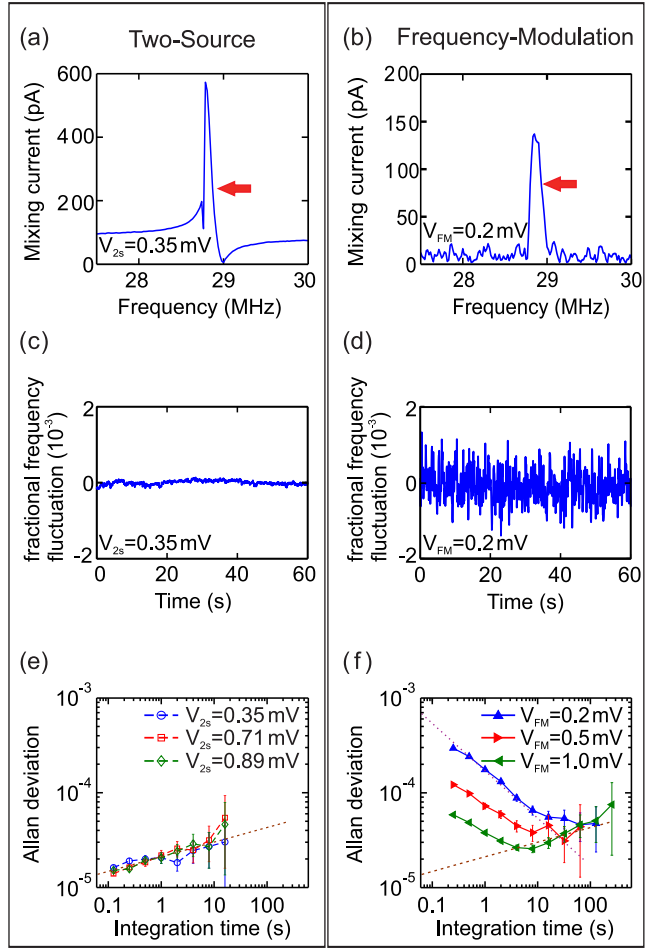}
\caption{\label{fig:fig_2} (a) Response of the nanotube device 1 to the driving frequency at 4.2\,K using the two-source setup. The large off-resonance current has a purely electrical origin \cite{sazonova_tunable_2004}. (b) Response of the nanotube resonator to the driving frequency at 4.2\,K using the FM setup. The integration time of the lock-in amplifier is 100\,ms. The red arrows in (a) and (b) indicate the slope used to measure the fluctuations of the resonance frequency. (c,d) Fractional frequency fluctuation $\delta\!f_{0}/f_{0}$ of the resonator measured with the two-source setup and the FM setup. Here $\delta\!f_{0}$ is the measured deviation of the resonance frequency at time $t$ with respect to the average resonance frequency $f_{0}$. (e) The frequency stability of the mechanical resonator measured with the two-source setup for different drives. The brown dashed line is a guide to the eye showing the trend of the Allan deviation as a function of the integration time. (f) The frequency stability of the mechanical resonator measured with the FM setup for different drives. The data at low integration time are compared to Eq. (\ref{eq:Allan_add_phase_noise}) (purple dotted line). The brown dashed line indicates the trend of the Allan deviation measured with the two-source setup in (e).}
\end{figure}

The Allan deviation measured with our two-source method is significantly better than the Allan deviation measured with FM (Figs. \ref{fig:fig_2}(a-f)). Both measurements were carried out in the same cool down. The best Allan deviation with the two-source is achieved at short integration times $\tau_{\mathrm{A}}$. This is of great interest for mass sensing and surface science experiments when adsorption, desorption, and diffusion processes are rapid. The Allan deviation with the two-source is independent of $S$, that is, the voltage amplitude $V_{\mathrm{2s}}$ applied to the gate. This indicates that the Allan deviation is limited by the noise of the resonance frequency of the nanotube resonator. The imprecision noise of the detection is irrelevant, since the corresponding Allan deviation is expected to be about two orders of magnitude smaller than the measured Allan deviation. The expected Allan deviation is 5.2$\cdot10^{-7}$ for an integration time $\tau_{\mathrm{A}}=1\,$s using Eq. (\ref{eq:Allan_add_phase_noise}) with $N_{\mathrm{T}}$=0.43\,pA/$\sqrt{\mathrm{Hz}}$.

By contrast, the Allan deviation measured with FM is given by the imprecision noise of the detection at low $\tau_{\mathrm{A}}$. The Allan deviation scales as $1/\sqrt{\tau_{\mathrm{A}}}$ and gets larger for lower $S$, as expected from Eq. (\ref{eq:Allan_add_phase_noise}). The measured Allan deviation is consistent with what is expected from the imprecision noise $N_{\mathrm{T}}$, since we obtain $N_{\mathrm{T}}= $18\,pA$/\sqrt{\mathrm{Hz}}$ from the measured $\sigma_{\mathrm{Allan}}(\tau_{\mathrm{A}})$ (dotted line in Fig. \ref{fig:fig_2}(f)) and Eq. (\ref{eq:Allan_add_phase_noise}), which is comparable to $N_{\mathrm{T}}= 16$\,pA$/\sqrt{\mathrm{Hz}}$ estimated from the off-resonance current fluctuations observed in Fig. \ref{fig:fig_2}(b). At long $\tau_{\mathrm{A}}$, the Allan deviation becomes similar to the Allan deviation measured with the two-source (Figs. \ref{fig:fig_2}(e,f)). This indicates that the Allan deviation becomes limited by the noise of the resonance frequency of the nanotube resonator.

\begin{figure}
\includegraphics{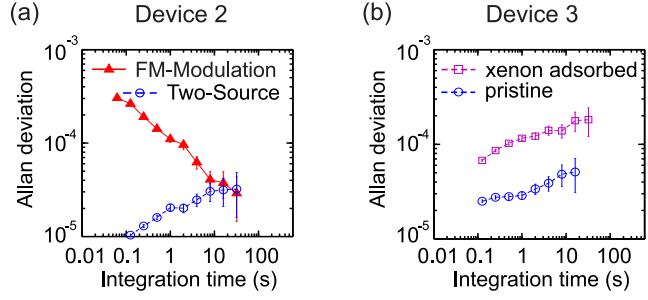}
\caption{\label{fig:fig_3} (a) The frequency stability of device 2 measured with the FM setup and the two-source setup. The FM measurement was performed with V$_\mathrm{FM}$=0.2\,mV and the two-source measurement with V$_{\mathrm{2s}}$=0.007\,mV. (b) The frequency stability of device 3 with and without xenon atoms adsorbed onto the nanotube. Both measurements were performed with V$_{\mathrm{2s}}$=0.071\,mV.}
\end{figure}

The $\sigma_{\mathrm{Allan}}(\tau_{\mathrm{A}})$ curves measured on three different devices with the two-source are similar (Figs. \ref{fig:fig_2}(e), \ref{fig:fig_3}(a,b)). This indicates that the physical origin of the noise of the resonance frequency is the same for the three devices. The Allan deviation reaches $\sim10^{-5}$ at 125\,ms integration time. Allan deviation measurements are usually compared to power law dependences, $\sigma_{\mathrm{Allan}} \propto \tau_{\mathrm{A}}^{\alpha}$. In our case, the exponent $\alpha$ approaches $\sim 1/4$. The positive slope of $\sigma_{\mathrm{Allan}}(\tau_{\mathrm{A}})$ plotted in a doubly-logarithm scale is often attributed to the drift of the resonance frequency due to the slow variations of the temperature and the voltage applied to the device. We characterized the fluctuations of the temperature, the static voltage applied to the gate electrode, and the amplitude of the high-frequency voltages applied to the device. These fluctuations correspond to Allan deviations that are between one and three orders of magnitude smaller than that measured in our devices (supplementary material). Therefore, the origin of the Allan deviation is not related to the drift of the temperature and the applied voltages.

Figure \ref{fig:fig_3}(b) shows how the Allan deviation is modified after having adsorbed a small number of xenon atoms onto the nanotube. The xenon atoms were administered through a small nozzle into the ultra-high vacuum cryostat. When impinging on the nanotube they have a certain sticking probability due to unspecific physisorption. From the shift of $f_0$, the number of xenon atoms is estimated to be 1.0 \% of the total number of carbon atoms in the suspended portion of the nanotube \cite{wang_phase_2010,tavernarakis_atomic_2014}. The presence of these xenon atoms significantly deteriorates the frequency stability of the device. The Allan deviation increases by a factor $\sim$~3 over the whole range of integration time. The deterioration of the frequency stability is attributed to the diffusion of xenon atoms over the surface of the nanotube, as reported in Ref.\,\cite{yang_surface_2011}. The reduced frequency stability is not related to adsorption/desorption processes, since the Allan deviation does not return to its initial value before the exposure to xenon while the measured pressure of the chamber does.

The slope of $\sigma_{\mathrm{Allan}}(\tau_{\mathrm{A}})$ plotted in the doubly-logarithm scale is positive with or without the adsorption of xenon atoms. This suggests that the frequency stability of our devices without adsorbed xenon in Figs. \ref{fig:fig_2}(e) and \ref{fig:fig_3}(a,b) is also limited by the diffusion of adsorbed atoms and molecules. These particles might come from the rest gas in our ultra-high vacuum setup, such as H$_2$, H$_2$O, CO, and CO$_2$. Our measurements are somewhat consistent with the model based on the diffusion of non-interacting particles, which predicts a positive slope. The noise due to diffusion is discussed in detail in the supplemental information of Ref.\,\cite{yang_surface_2011}. The typical exponent measured in our work is about half the value expected from this simple model where trapping of particles at defect sites and particle-particle interaction are both disregarded. A more complete characterization of the physics of the frequency stability of nanotube resonators is beyond the scope of the Letter and will require further work in the future.

In conclusion, we developed a method to measure the frequency stability of nanotube mechanical resonators with an unprecedented quality. The frequency noise measured at 4.2\,K is limited by the resonator itself, and not by the imprecision of the measurement. The origin of the frequency noise might be related to the diffusion of particles over the surface of the nanotube. Our detection method holds promise for studying the diffusion of atoms and molecules over crystalline surfaces \cite{yang_surface_2011}, the interplay between particle diffusion and mechanical vibrations \cite{atalaya_diffusion-induced_2011,rhen_particle_2016}, and the phase transition in monolayers of adsorbed atoms \cite{wang_phase_2010,tavernarakis_atomic_2014}.

See supplementary material for the quantification of the different instrumentation related noise sources and temperature fluctuations.

The authors acknowledge financial support from the ERC advanced grant 692876, the Foundation Cellex, the CERCA Programme, AGAUR, Severo Ochoa (SEV-2015-0522), the grant FIS2015-69831-P of MINECO, and the Fondo Europeo de Desarrollo Regional (FEDER). J. S. acknowledges the financial support by the la Caixa-Severo Ochoa PhD fellowship.
We wish to thank Mar\'ia Jos\'e Esplandiu Egido for support in the nanotube growth and Brian Thibeault for help in fabrication.

\bibliography{references}

\end{document}